\documentclass[pra,superscriptaddress,showpacs,twocolumn,10pt]{revtex4}
\usepackage{amssymb}
\usepackage{amsmath}
\usepackage{graphicx}

\begin{document}%

\newcommand{\ket}[1]{|#1\rangle}
\newcommand{\bra}[1]{\langle#1|}
\newcommand{\inner}[2]{\langle#1|#2\rangle}
\newcommand{\lr}\longrightarrow
\newcommand{\ra}\rightarrow
\newcommand{\tr}{{\rm Tr}}
\newcommand{\per}{{\rm per}}
\newcommand{\sgn}{{\rm sgn}}
\newcommand{\fsp}{{\rm span}}
\newcommand{\fsup}{{\rm supp}}
\newcommand{\fdg}{{\rm diag}}

\newtheorem{thm}{Theorem}
\newtheorem{Prop}{Proposition}
\newtheorem{Coro}{Corollary}
\newtheorem{Lemma}{Lemma}
\newtheorem{Def}{Definition}

\title{Discrimination between pure states and mixed states}
\author{Chi Zhang}
\email{zangcy00@mails.tsinghua.edu.cn}
\author{Guoming Wang}
\email{wgm00@mails.tsinghua.edu.cn}
\author{Mingsheng Ying}
\email{yingmsh@tsinghua.edu.cn} \affiliation{State Key Laboratory
of Intelligent Technology and Systems, Department of Computer
Science and Technology Tsinghua University, Beijing, China,
100084}
\date{\today}

\begin{abstract}
In this paper, we discuss the problem of determining whether a
quantum system is in a pure state, or in a mixed state. We apply two
strategies to settle this problem: the unambiguous discrimination
and the maximum confidence discrimination. We also proved that the
optimal versions of both strategies are equivalent. The efficiency
of the discrimination is also analyzed. This scheme also provides a
method to estimate purity of quantum states, and Schmidt numbers of
composed systems.
\end{abstract}

\maketitle

\section{Introduction}

In many applications of quantum information, one of the important
elements which affect the result of quantum process, is the purity
of the quantum states produced or utilized. Hence, an interesting
and important problem in quantum information is to estimate the
purity of a quantum system \cite{E1,E2,E3,E4,E5}. This problem is
also strongly related to the estimation of the entanglement of
multiparty systems \cite{EE1,EE2,EE3}.

However, all above references considered this problem only in the
simplest case of qubits, where both the definition and computation
about purity are clear. Estimating the purity of a general quantum
system is still open. In this paper, we first consider an extreme
situation: given some copies of a quantum state, the task of us is
to determine whether the state is pure or mixed. The process is
called discrimination between pure states and mixed states. Then, by
counting different results obtained in the above discriminations, we
offer an effective method to estimate the purity of quantum states.
The idea of discrimination between pure states and mixed states, was
first mentioned in Ref.\cite{EE3}. However, they did not study the
problem formally and systematically, which is our aim in this paper.
There are two different strategies to design the discrimination: the
unambiguous discrimination \cite{unambiguous}, and the maximum
confidence discrimination \cite{confidence}. In the strategy of
unambiguous discrimination, one can tell whether the quantum system
is in a pure state or a mixed state without error, but a non-zero
probability of inconclusive answer is allowed. In this paper, in
order to simplify the presentation, we use the term ``unambiguous''
in a more general sense: it allows the success probability to be
zero in some situation. On the other hand, in the maximum confidence
discrimination, an inconclusive answer is not allowed, and after
each discrimination, we must give a statement whether the quantum
state is pure, or mixed. The discrimination is so named, because if
an answer is given, the probability of obtaining a correct
conclusion is maximized.

It is convenient to introduce some notations here. In the Hilbert
space $H^{\otimes n}$, we use $H^{\otimes n}_{sym}$ to denote its
symmetric subspace \cite{book_space}. The orthogonal complement of
$H^{\otimes n}_{sym}$ is called the asymmetric subspace of
$H^{\otimes n}$, and denoted as $H^{\otimes n}_{asym}$. We use
$\Phi(H^{\otimes n}_{sym})$ and $\Phi(H^{\otimes n}_{asym})$ to
represent the projectors of these two subspaces respectively. In
this paper, we prove that, given $n$ copies of a quantum state
$\rho$ in Hilbert space $H$, the optimal unambiguous discrimination
and the maximum confidence discrimination can be carried out by the
same measurement $\{\Pi_0=\Phi(H^{\otimes
n}_{sym}),\Pi_1=\Phi(H^{\otimes n}_{asym})\}$. The difference
between these two discriminations comes only from the different
explanations of the outcomes. In the unambiguous discrimination, the
outcome `$0$' is an inconclusive answer, and the outcome `$1$'
indicates that the system is in a mixed state. The drawback of the
unambiguous discrimination is that, if the quantum system is in a
pure state, people always fail to give a confirm answer. However, in
the maximum confidence discrimination, the outcome `$0$' indicates
that the quantum system is considered to be in a pure state, and the
outcome `$1$' indicates a mixed state.

There are two natural assumptions in this paper. First, the purity
of quantum states is invariant under any unitary operation. Suppose
the purity of a quantum state $\rho$ is represented by $\mu(\rho)$,
it must satisfy that $\mu(\rho) = \mu(U\rho U^\dagger)$, for any
unitary operator $U$. We also assume that, when $\rho$ is a pure
state, $\mu(\rho)=1$, otherwise $0\leq\mu(\rho)<1$. For instance,
the usually used purity of quantum states, $\mu(\rho) =
\tr^2(\rho)$, clearly satisfies these conditions. Second, the priori
probability distributions of quantum states are also assumed
invariant under unitary operations. Let us denote the priori
probability density function as $\eta(\rho)$, then $\eta(\rho) =
\eta(U\rho U^\dagger)$, for any unitary operator $U$.

Our present article is organized as follows. In section.\ref{s2}, we
provide the optimal unambiguous discrimination between pure states
and mixed states. And, in section.\ref{mc}, we provide the maximum
confidence discrimination between pure states and mixed states. We
also generalize the unambiguous discrimination between pure states
and mixed states to a ``semi-unambiguous'' estimation for ranks of
quantum states in section.\ref{s_rank}, which can also be seen as an
estimation for the Schmidt number of bipartite quantum systems.
Finally, in section.\ref{ep}, we provide a strategy to estimate the
purity of quantum states.

\section{Optimal Unambiguous Discrimination}\label{s2}

In this section, we consider the unambiguous discrimination between
pure states and mixed states. Suppose we are given $n$ copies of a
quantum state, which is in the Hilbert space $H$. The unambiguous
discrimination is described by a POVM measurement on the Hilbert
space $H^{\otimes n}$. The measurement is comprised by three
positive operators, $\Pi_p$, $\Pi_m$, and $\Pi_?$, satisfying that
\begin{equation}\label{cop}
\tr(\Pi_p \rho^{\otimes n}) = 0,
\end{equation}
for any mixed state $\rho$,
\begin{equation}\label{com}
\bra{\psi}^{\otimes n} \Pi_m \ket{\psi}^{\otimes n} = 0,
\end{equation}
for any pure state $\ket{\psi}$, and
\begin{equation}
\Pi_? = I - \Pi_m - \Pi_p.
\end{equation}
Therefore, if the outcome is `p', the system is assured to be in a
pure state; if the outcome is `m', the system is in a mixed state;
and outcome `?' denotes an inconclusive answer.

The efficiency of the discrimination is
\begin{equation}
\begin{split}
p &= \int_{0\leq\mu(\rho)<1} \tr(\rho^{\otimes n}\Pi_m) \eta(\rho)
d\rho\\ &+ \int_{\mu(\rho)=1} \tr(\rho^{\otimes n}\Pi_p)
\eta(\rho) d\rho.
\end{split}
\end{equation}
The optimal unambiguous discrimination is the one with the maximum
efficiency, and we have the following theorem.

\begin{thm}\label{thm1}
The optimal unambiguous discrimination between pure states and mixed
states is a POVM measurement $\{\Pi_p, \Pi_m, \Pi_?\}$, such that
\begin{equation}\label{un}
\begin{split}
\Pi_p &= 0,\\ \Pi_m &= \Phi(H^{\otimes n}_{asym}),\\ \Pi_? &=
\Phi(H^{\otimes n}_{sym}),
\end{split}
\end{equation}
where $\Phi(H^{\otimes n}_{sym})$ and $\Phi(H^{\otimes n}_{asym})$
are the projectors of symmetric subspace and asymmetric subspace of
$H^{\otimes n}$ respectively.
\end{thm}

{\it Proof.} For a mixed state $\rho$, whose spectrum decomposition
is $\rho = \sum_{i=1}^{m} \lambda_i \ket{\phi_i}\bra{\phi_i}$, and
for any $n$-tuple chosen from $\{1,\cdots,m\}$,
$\pi=(\pi_1,\cdots,\pi_n)$, where repetition is allowed, let us
introduce the following two definitions,
\begin{equation}
\lambda_{\pi} = \prod_{j=1}^{n} \lambda_{\pi_j},
\end{equation}
and
\begin{equation}
\ket{\phi_\pi} = \otimes_{j=1}^n \ket{\phi_{\pi_j}}.
\end{equation}
Then,
\begin{equation}\label{mix}
\rho^{\otimes n} = \sum_{\pi} \lambda_{\pi}
\ket{\phi_{\pi}}\bra{\phi_{\pi}},
\end{equation}
where $\pi$ ranges over all $n$-tuples chosen from $\{1,\cdots,m\}$.

Because $\Pi_p$ is a positive operator, from Eq.(\ref{mix}) and
Eq.(\ref{cop}),
\begin{equation}
\bra{\phi_\sigma}\Pi_p\ket{\phi_\sigma} = 0,
\end{equation}
for any product state $\ket{\phi_\sigma}$. Therefore, when the
system is in a pure state $\ket{\psi}$, it satisfies that
\begin{equation}
\bra{\psi}^{\otimes n} \Pi_p \ket{\psi}^{\otimes n} = 0,
\end{equation}
i.e., for any situation, the probability of getting the `p' result
is always zero, which means that without loss of generality, we can
simply let $\Pi_p = 0$.

Then, from Eq.(\ref{com}), we know that $\Pi_m$ is orthogonal to any
$\ket{\psi}^{\otimes n}$, where $\ket{\psi}\in H$. It is known that
the span space of all $\ket{\psi}^{\otimes n}$ is just the symmetric
subspace of $H^{\otimes n}$, which has been denoted as $H^{\otimes
n}_{sym}$ \cite{comparison}. Thus, the support space of $\Pi_m$ must
be in the asymmetric subspace $H^{\otimes n}_{asym}$, i.e., $\Pi_m
\leq \Phi(H^{\otimes n}_{asym})$. The probability of determining a
mixed state $\rho$ is
\begin{equation}
p(m|\rho) = \tr(\rho^{\otimes n}\Pi_m) \leq \tr(\rho^{\otimes n}
\Phi(H^{\otimes n}_{asym})).
\end{equation}
Hence, the optimal unambiguous discrimination is the measurement
$\{\Pi_p, \Pi_m,\Pi_?\}$ given in Eq.(\ref{un}). \hfill $\Box$

Under the optimal unambiguous discrimination, when the quantum
system is in a pure state, the result is sure to be inconclusive. If
the quantum system is in a mixed state $\rho$, the probability of
receiving an inconclusive answer is
\begin{equation}\label{1_1}
\begin{split}
p(?|\rho) &= \tr(\rho^{\otimes n}\Pi_?)\\
&= \sum_{\pi} \lambda_{\pi}\bra{\phi_{\pi}}\Phi(H^{\otimes n}_{sym})\ket{\phi_{\pi}}\\
&= \sum_{\pi} \frac{\lambda_{\pi}}{n!} \per(\Gamma_{\pi}),
\end{split}
\end{equation}
where $\Gamma_{\pi}$ is the Gram matrix derived from
$\{\ket{\phi_{\pi_1}},\cdots,\ket{\phi_{\pi_n}}\}$, and $\per(A)$
denotes the permanent of the matrix $A$, i.e.,
\begin{equation}
\per(A) = \sum_{\sigma}\prod_{i} A(i,\sigma(i)),
\end{equation}
where $\sigma$ ranges over all permutation on $n$ symbols
\cite{book_space}.

Let $\pi$ is an $n$-tuple valued in $\{1,\cdots,m\}$. We use
$n^{\pi}_i$ to denote the number of occurrences of $i$ in $\pi$,
where $i=1,\cdots,m$. Because for any two eigenvectors of $\rho$
with non-zero eigenvalues, $\inner{\phi_i}{\phi_j} = \delta_{i,j}$.
\begin{equation}
\Gamma_{\pi} = \bigoplus_{i=1}^m I_{n^{\pi}_i},
\end{equation}
where $I_{n^{\pi}_i}$ is the $n^{\pi}_i$-dimensional identity
matrix. Consequently, from Eq.(\ref{1_1})
\begin{equation}\label{1_2}
\begin{split}
p(?|\rho) &= \sum_{\pi} \frac{\lambda_{\pi}}{n!} \prod_{i=1}^m
n^{\pi}_i!\\
&= \sum_{\sum_{i=1}^m n_i=n}\frac{n!}{\prod_{i=1}^m n_i!}
\prod_{i=1}^m \lambda_{i}^{n_i} \frac{\prod_{i=1}^m n_i!}{n!}\\
&= \sum_{\sum_{i=1}^m n_i=n} \prod_{i=1}^m \lambda_{i}^{n_i}.
\end{split}
\end{equation}

\section{Semi-unambiguous Estimation of Schmidt number}\label{s_rank}

As we know, the entanglement of a bipartite quantum system is
closely related to the purity of one of its subsystems. Whether a
subsystem is in a pure state is equivalent to whether the total
quantum system is in a product state. Hence, the measurement given
in the above section also provides an unambiguous estimation for
entanglement of bipartite quantum systems. Moreover, in this
section, we will provide a natural generalization, which can be
called semi-unambiguous estimation of the Schmidt number of
bipartite systems.

The Schmidt number of a bipartite system equals to the rank of the
quantum state in each of its subsystems. Hence, estimating the
Schmidt number is equivalent to estimating the rank of quantum
states. First, let us reconsider the discrimination between pure
states and mixed states. In the discrimination, the `m' result means
that the rank of the state is no less than $2$, while the
inconclusive answer can also be considered as a trivial conclusion
that the rank of the state is no less than $1$. Although the
discrimination does not offer the exact value of the rank of the
quantum state, it offers a lower bound for the rank. Moreover, the
lower bound is assured to be correct. In this mean, we can call the
discrimination between pure states and mixed states also a
``semi-unambiguous'' estimation for the rank of quantum states. A
more general ``semi-unambiguous'' estimation of the rank of quantum
states can be defined as a POVM measurement on $H^{\otimes n}$ with
operators $\{\Pi_1,\Pi_2,\cdots,\Pi_m\}$, where $m$ is the dimension
of $H$. The measurement satisfies that for any quantum state $\rho$
whose rank is $k$, $\tr(\Pi_i\rho^{\otimes n}) = 0$, for any $i>k$.
Thus, whenever the outcome $k$ is observed, we can make sure that
the rank of $\rho$ is no less than $k$.

Before providing the semi-unambiguous estimation of the rank of
quantum states. We first introduce some fundamental knowledge about
group representation theory needed here. For details, please see
Ref. \cite{book2}.

A Young diagram $[\lambda] = [\lambda_1,\cdots,\lambda_k]$, where
$\sum \lambda_i = n$ and
$\lambda_1\geq\lambda_2\geq\cdots\lambda_k>0$, is a graphical
representation of a partition of a natural number $n$. It
consists of $n$ cells, arranged in left-justified rows, where the
number of cells in the $i$th row is $\lambda_i$.

A Young tableau is obtained by placing the numbers $1,\cdots,n$
in the $n$ cells of a Young diagram. If the numbers form an
increasing sequence along each row and each column, the Young
tableau is called standard Young tableau. For a given Young
diagram $[\lambda]$, the number of standard Young tableau can be
calculated with the hook length formula, and denoted by
$f^{[\lambda]}$. In this paper, we use $T^{[\lambda]}_r$ to
denoted the $r$th standard Young tableau, where
$r=1,\cdots,f^{[\lambda]}$.

The Hilbert space $H^{\otimes n}$, where the dimension of $H$ is
$m$, can be decomposed into a set of invariant subspaces under
operation $U^{\otimes n}$, for any unitary operation $U$ on $H$.
Each of the subspaces corresponds to a standard Young tableau
$T^{[\lambda]}_r$, where the number of rows in $[\lambda]$ is no
more than $m$. So, we can denote the subspaces as $H^{[\lambda]}_r$,
and denote its projector as $\Phi(H^{[\lambda]}_r)$. Then, we have
\begin{equation}
H^{\otimes n} = \bigoplus_{[\lambda],r} H^{[\lambda]}_{r}.
\end{equation}
For instance, the symmetric subspace $H^{\otimes n}_{sym}$ is just
one of these subspaces, $H^{\square\square\cdots\square}_1$.

For a quantum state $\rho$ in $H$, whose rank is $k$, the support
space of $\rho^{\otimes n}$ is in the sum of subspaces
$H^{[\lambda]}_r$, where the number of rows in Young diagram
$[\lambda]$ is no greater than $k$. Therefore, a semi-unambiguous
estimation of the rank of quantum states can be designed as a POVM
measurement $\{\Pi_1,\cdots,\Pi_m\}$, such that
\begin{equation}\label{de}
\Pi_i = \sum_{h([\lambda])=i}\sum_{r}\Phi(H^{[\lambda]}_{r}),
\end{equation}
where $h([\lambda])$ is the number of rows in $[\lambda]$. As said
above, if the rank of $\rho$ is $k$, $\tr(\Pi_i\rho^{\otimes n}) =
0$, for any $i>k$. Thus, once an `$i$' result is observed, we can
assert that the rank of $\rho$ is no less than $i$. For $n$ copies
of a bipartite quantum system, through measuring any of its
subsystems with the measurement given in Eq.(\ref{de}), we can
semi-unambiguously estimate the Schmidt number of the whole system.

\section{Maximum Confidence Discrimination}\label{mc}

In this section, we consider a different strategy for determining
whether the quantum system is in a pure state, which is called
``maximum confidence discrimination''\cite{confidence}.

The discrimination is still a POVM measurement $\{\Pi_p, \Pi_m\}$.
But when the outcome is `m', the quantum system is believed in a
mixed state; otherwise, the outcome is `p', and the quantum state is
considered to be pure. A maximum confidence strategy is to maximize
the reliability of the conclusion, i.e., let the following two
probabilities be maximized,
\begin{equation}\label{2_1}
p(pure|p) = \frac{\int_{\mu(\rho)=1} \eta(\rho)\tr(\rho^{\otimes
n}\Pi_p) d\rho}{\int_{\mu(\rho)\leq 1} \eta(\rho)\tr(\rho^{\otimes
n}\Pi_p) d\rho},
\end{equation}
and
\begin{equation}\label{2_2}
p(mixed|m) = \frac{\int_{\mu(\rho)<1} \eta(\rho)\tr(\rho^{\otimes
n}\Pi_m) d\rho}{\int_{\mu(\rho)\leq 1} \eta(\rho)\tr(\rho^{\otimes
n}\Pi_m)d\rho}.
\end{equation}

Clearly, Eq.(\ref{2_1}) and Eq.(\ref{2_2}) do not always get maximum
values at the same time. However, on the assumptions about unitary
invariance of $\eta(\rho)$ and $\mu(\rho)$, we can prove that there
exists a measurement $\{\Pi_p,\Pi_m\}$ maximizing both
Eq.(\ref{2_1}) and Eq.(\ref{2_2}), as the following theorem states.

\begin{thm}\label{thm2}
The maximum confidence discrimination between pure states and
mixed states is a POVM measurement $\{\Pi_p, \Pi_m\}$, such that
\begin{equation}
\begin{split}
\Pi_p &= \Phi(H^{\otimes n}_{sym}),\\
\Pi_m &= \Phi(H^{\otimes n}_{asym}),
\end{split}
\end{equation}
where $\Phi(H^{\otimes n}_{sym})$ and $\Phi(H^{\otimes n}_{asym})$
are as in Theorem \ref{thm1}.
\end{thm}

{\it Proof.} First, we consider the construction of $\Pi_p$. From
the assumptions that $\eta(\rho)=\eta(U\rho U^{\dagger})$ and
$\mu(\rho)=\mu(U\rho U^{\dagger})$,
\begin{equation}
\begin{split}
p(pure|p) &= \frac{\int_{\mu(\rho)=1} \eta(\rho)\tr(\rho^{\otimes
n}\Pi_p) d\rho}{\int_{\mu(\rho)\leq 1} \eta(\rho)\tr(\rho^{\otimes
n}\Pi_p) d\rho}\\ &= \frac{\int_{\mu(\rho)=1}
\eta(\rho)\tr(\rho^{\otimes n}U^{\otimes n}\Pi_p
(U^{\dagger})^{\otimes n}) d\rho}{\int_{\mu(\rho)\leq 1}
\eta(\rho)\tr(\rho^{\otimes n}U^{\otimes n}\Pi_p
(U^{\dagger})^{\otimes n}) d\rho},
\end{split}
\end{equation}
for any unitary operation $U$. Hence, if $\Pi_p$ maximizes
Eq.(\ref{2_1}), so does $\int U^{\otimes
n}\Pi_p(U^{\dagger})^{\otimes n} dU$ with respect to the normalized
invariant measure $dU$ of the unitary group $U(m)$. Hence, we can
choose the operator $\Pi_p$ to satisfy that
\begin{equation}
\Pi_p = \int U^{\otimes n}\Pi_p(U^{\dagger})^{\otimes n} dU,
\end{equation}
which shows that $\Pi_p$ commutes with any unitary operator of the
form $U^{\otimes n}$. Thus, from the representation theory of
classical groups in Ref.{\cite{book3}}, $\Pi_p$ can be expressed as
a linear combination of permutation operators
\begin{equation}\label{struct}
\Pi_p = \sum_{\sigma} \alpha_{\sigma}V_{\sigma},
\end{equation}
where $\alpha_{\sigma}\in C$, $\sigma$ ranges over all permutations
of $n$ elements, and $V_{\sigma}$ is the permutation operator
derived from $\sigma$, i.e.,
\begin{equation}
V_{\sigma} \ket{\psi_1}\ket{\psi_2}\cdots\ket{\psi_n} =
\ket{\psi_{\sigma_1}}\ket{\psi_{\sigma_2}}\cdots
\ket{\psi_{\sigma_n}}.
\end{equation}

For any state $\ket{\varphi}$ in the symmetric subspace $H^{\otimes
n}_{sym}$, $V_{\sigma}\ket{\varphi} = \ket{\varphi}$, so
$\Pi_p\ket{\varphi}=(\sum_{\sigma} \alpha_{\sigma})\ket{\varphi}$,
which indicates that
\begin{equation}
\Pi_p = \alpha\Phi(H^{\otimes n}_{sym}) \oplus \Pi'_p,
\end{equation}
where $\Pi'_p$ is a positive operator whose support space is in
$H^{\otimes n}_{asym}$, and $\alpha = \sum_{\sigma}
\alpha_{\sigma}$. Because for any pure state
$\rho=\ket{\psi}\bra{\psi}$, the support space of $\rho^{\otimes n}$
is in the symmetric subspace $H^{\otimes n}_{sym}$,
$\tr(\rho^{\otimes n}\Pi'_p)=0$ for any $\mu(\rho)=1$. Therefore,
the numerator of Eq.(\ref{2_1}) does not change if we substitute
$\Pi_p$ with $ \alpha\Phi(H^{\otimes n}_{sym})$, and the denominator
diminishes or remains the same. So, the optimal $\Pi_p$ has the form
of $\alpha\Phi(H^{\otimes n}_{sym})$ for any constant $\alpha$.

On the other hand, if we choose $\Phi(H^{\otimes n}_{asym})$ as
$\Pi_m$, then for any pure state $\rho$, whose purity $\mu(\rho)=1$,
we have $\tr(\rho^{\otimes n}\Pi_m) = 0$, and Eq.(\ref{2_2}) has the
maximum value $1$. To satisfy the condition $\Pi_p + \Pi_m = I$, let
$\alpha=1$, $\Pi_p = \Phi(H^{\otimes n}_{sym})$. This completes the
proof. \hfill $\Box$

It is easy to see that the optimal unambiguous discrimination and
the maximum confidence discrimination are the same measurement
$\{\Pi_0 = \Phi(H^{\otimes n}_{sym}), \Pi_1 = \Phi(H^{\otimes
n}_{asym})\}$. The difference between the two discriminations is the
meaning of the `$0$' result. In the former discrimination, the `$0$'
result means an inconclusive answer; however, in the latter
discrimination, if a `$0$' result is obtained, the quantum system is
considered in a pure state.

From Eq.(\ref{1_2}), for $n$ copies of a quantum state $\rho$, under
the measurement of $\{\Pi_0,\Pi_1\}$ given above, the probability of
receiving a `$0$' result is
\begin{equation}
p_0(n) = \sum_{\sum_{i=1}^m n_i=n} \prod_{i=1}^m
\lambda_{i}^{n_i},
\end{equation}
where $\lambda_1,\cdots,\lambda_m$ are the eigenvalues of $\rho$. As
we know, the above quantity is the complete symmetric polynomial of
degree $n$ for $\{\lambda_1,\cdots,\lambda_m\}$, which is usually
denoted by $h_n(\lambda_1,\cdots,\lambda_m)$. From
Ref.\cite{symmetric polinomial}, the complete symmetric polynomials
can be derived from a generating function
\begin{equation}\label{2_3}
H_m(t) = \sum_{k\geq 0}h_k(\lambda_1,\cdots,\lambda_m)t^{k}
=\frac{1}{\prod_{i=1}^m (1-t\lambda_i)}.
\end{equation}

Let $\lambda^*$ stand for the maximum eigenvalue of $\rho$, then, if
we have $n$ copies of the states, the probability of judging it to
be pure can be evaluated as follows:
\begin{equation}
\begin{split}
p_0(n) &= \sum_{\sum_{i=1}^m n_i=n} \prod_{i=1}^m
\lambda_{i}^{n_i}\\ &\leq  \dbinom{n+m-1}{n}(\lambda^*)^n.
\end{split}
\end{equation}
Then, if the quantum system is in a pure state, $p_0(n)$ will always
be $1$, otherwise $\lambda^* < 1$, and $p_0(n)$ will converge to
zero with exponential convergence rate.

In section.\ref{s_rank}, we discuss the semi-unambiguous estimation
of the rank of quantum states, which is given in Eq.(\ref{de}). An
open problem is whether this measurement also offers a maximum
confidence estimation of ranks of quantum states, if we consider the
result `$i$' as a claim that the rank of the quantum state is $i$.

\section{Estimating purity of states}\label{ep}

The maximum confidence discrimination between pure states and mixed
states provides a natural intuition for the purity of a quantum
system, i.e., the greater the probability of getting a `0' result,
the closer it is to a pure state. Hence, by repetitively performing
the measurement, and counting the proportion of `0' results, we can
estimate the probability of judging the system being pure, which, in
some sense, reflects some information about the purity of the
system. However, a more interesting conclusion is that, no matter
how people define the purity of quantum states, as long as it
satisfies the condition of unitary invariant, it can be well
estimated through a set of maximum confidence discriminations.

On the assumption of unitary invariant, the purity of a quantum
state $\rho$, $\mu(\rho) = \mu(\fdg(\lambda_1,\cdots,\lambda_m))$,
where $\lambda_1,\cdots,\lambda_m$ are the eigenvalues of $\rho$.
Hence, $\mu(\rho)$ is a function of its eigenvalues. Estimating the
purity of a quantum state $\rho$ can be reduced to estimating the
eigenvalues of $\rho$. The characteristic polynomial of $\rho$ is a
polynomial, whose roots are the eigenvalues, i.e.,
\begin{equation}\label{3_3}
\det(xI-\rho) = \prod_{i=1}^m (x-\lambda_i) = \sum_{j=0}^m a_j
x^{m-j}.
\end{equation}
If we can successfully estimate every coefficient $a_j$,
$j=0,\cdots,m$, the eigenvalues can be estimated by solving the
equation $\sum_{j=0}^m a_j x^{m-j} = 0$.

Recall the famous Viete's theorem, it is easy to know
\begin{equation}\label{3_0}
\begin{split}
a_0 &= e_0(\lambda_1,\cdots,\lambda_m) =1 \\
a_1 &= -e_1(\lambda_1,\cdots,\lambda_m) = -\sum_{i=1}^{m}
\lambda_i\\
a_2 &= e_2(\lambda_1,\cdots,\lambda_m) = \sum_{1\leq i_1<i_2\leq
n} \lambda_{i_1}\lambda_{i_2}\\
\cdots\\
a_k &= (-1)^{k}e_k(\lambda_1,\cdots,\lambda_m) =
(-1)^{k}\sum_{1\leq
i_1<\cdots<i_k\leq n}\lambda_{i_1}\cdots\lambda_{i_k}\\
\cdots\\
a_m &= (-1)^m e_m(\lambda_1,\cdots,\lambda_m) = (-1)^{m}
\lambda_1\lambda_2\cdots\lambda_m.
\end{split}
\end{equation}
Here, the polynomial $e_k(\lambda_1,\cdots,\lambda_m)$ is the $m$-th
elementary symmetric polynomial of $\{\lambda_1,\cdots,\lambda_m\}$
\cite{symmetric polinomial}, whose generating function is,
\begin{equation}\label{3_1}
E_m(t) = \sum_{i=0}^{m} e_i(\lambda_1,\cdots,\lambda_m)t^{i} =
\prod_{i=1}^{m}(1+t\lambda_i).
\end{equation}
Combined with Eq.(\ref{2_3}), we have that $H(t)E(-t) = 1$, so
\begin{equation}
\sum_{r=0}^k (-1)^r e_r h_{m-r} =0,
\end{equation}
for any $k\geq 1$, if we set $e_r(\lambda_1,\cdots,\lambda_m)= 0$,
when $r>m$. Here, for simplicity, we use $e_k$, $h_l$ to denote
$e_k(\lambda_1,\cdots,\lambda_m)$, $h_l(\lambda_1,\cdots,\lambda_m)$
respectively. Then, it is not hard to see that
\begin{equation}\label{3_2}
e_k = \begin{vmatrix} h_1& h_2& h_3& \cdots & h_{k-1}& h_k\\
1& h_1& h_2& \cdots & h_{k-2}& h_{k-1}\\
0& 1& h_1& \cdots & h_{k-3}& h_{k-2}\\
\vdots & \vdots & \vdots & & \vdots & \vdots\\
0& 0& 0& \cdots & h_1 & h_2 \\
0& 0& 0& \cdots & 1 & h_2
\end{vmatrix},
\end{equation}

Clearly, $h_1=\sum_{i=1}^m \lambda_i =\tr(\rho) = 1$. As stated in
section.\ref{mc}, for any $k\geq 2$, $h_k$ is the probability of
receiving `$0$' result, when we measure $\rho^{\otimes k}$ by the
measurement $\{\Pi_0 = \Phi(H^{\otimes k}_{sym}), \Pi_1 =
\Phi(H^{\otimes k}_{asym})\}$. Therefore, if we have $N$ copies of
quantum state $\rho$, where $N$ is much larger than $m$, we can
estimate the eigenvalues of $\rho$ in the following strategy.

First, separate the $N$ copies into $m$ groups, the $k$th group has
$kN_k$ copies of the quantum state. Then, operate the measurement
$\{\Pi_0 = \Phi(H^{\otimes k}_{sym}), \Pi_1 = \Phi(H^{\otimes
k}_{asym})\}$ on $\rho^{\otimes k}$ for $N_k$ times in the $k$th
group. Suppose among these results, the number of `0' results is
$S_k$, then we can estimate $p_0(k)$, i.e., $h_k$ by
$\frac{S_k}{N_k}$. Then, through Eq.(\ref{3_2}), we can estimate
every $e_k$, where $1\leq k\leq m$. Hence, from Eq.(\ref{3_0}), the
characteristic polynomial of $\rho$, whose roots are the eigenvalues
we want to estimate is known. The task remained for us is to solve
the equation given in Eq.(\ref{3_3}).

\section{Conclusion}

In this paper, we investigate the discrimination between pure states
and mixed states, which may play an important role in further study
for estimating the purity of quantum states. The discrimination is
described by POVM measurements $\{\Pi_0 = \Phi(H^{\otimes n}_{sym}),
\Pi_1 = \Phi(H^{\otimes n}_{asym})\}$ on $n$ copies of the quantum
state being discriminated. If the `$0$' result is considered as an
inconclusive answer, the measurement is the optimal unambiguous
discrimination. On the other hand, if the `$0$' result is considered
as a hint that the quantum system is in a pure state, the
discrimination is the maximum confidence discrimination. We also
provide a semi-unambiguous estimation for the rank of quantum
states, which also can be used to estimate the Schmidt number of
bipartite quantum systems. Finally, we give a strategy to estimate
the purity of quantum systems.

\end{document}